\documentclass{pasj00}
\begin{document}
\SetRunningHead{T. Kato et al.}{Superhumps in HO Delphini}

\Received{}%{yyyy/mm/dd}
\Accepted{}%{yyyy/mm/dd}

\title{Superhumps in the Rarely Outbursting SU UMa-Type Dwarf Nova,\\ HO Delphini}

\author{Taichi \textsc{Kato}}
\affil{Department of Astronomy, Kyoto University,
       Sakyo-ku, Kyoto 606-8502}
\email{tkato@kusastro.kyoto-u.ac.jp}

\author{Daisaku \textsc{Nogami}}
\affil{Hida Observatory, Kyoto University, Kamitakara, Gifu 506-1314}
\email{nogami@kwasan.kyoto-u.ac.jp}

\author{Marko \textsc{Moilanen}}
\affil{Nyr\"{o}l\"{a} Observatory, Jyv\"{a}skyl\"{a}n Sirius ry, Kyllikinkatu
       1, FIN-40100 Jyv\"{a}skyl\"{a}, Finland}

\email{\rm{and}}

\author{Hitoshi \textsc{Yamaoka}}
\affil{Faculty of Science, Kyushu University, Fukuoka 810-8560}
\email{yamaoka@rc.kyushu-u.ac.jp}


\maketitle

\begin{abstract}
   We observed the 1994, 1996 and 2001 outbursts of HO Del.  From the
detection of secure superhumps, HO Del is confirmed to be an SU UMa-type
dwarf nova with a superhump period of 0.06453(6) d.  Based on the recent
observations and the past records, the outbursts of HO Del are found to
be relatively rare, with the shortest intervals of superoutbursts being
$\sim$740 d.  Among SU UMa-type dwarf novae with similar outburst
intervals, the outburst amplitude ($\sim$5.0 mag) is unusually small.
HO Del showed a rather rapid decay of the superhump amplitudes, and no
regrowth of the amplitudes during the later stage, in contrast to the
commonly observed behavior in SU UMa-type dwarf novae with long outburst
intervals.  We positively identified HO Del with a ROSAT X-ray
source, and obtained a relatively large X-ray luminosity of
10$^{31.1\pm0.2}$ erg s$^{-1}$.  We also performed a literature survey of
SU UMa-type dwarf novae, and presented a new set of basic statistics.
The SU UMa-type dwarf novae with a brightening trend or with a regrowth
of superhumps near the termination of a superoutburst are found to be
rather tightly confined in a small region on the
(superhump period--supercycle length) plane.  These features may provide
a better observational distinction for the previously claimed subgroup of
dwarf novae (Tremendous Outburst Amplitude Dwarf Novae).
\end{abstract}

\section{Introduction}

   Dwarf novae (DNe) are a class of cataclysmic variables (CVs), in which
the instabilities in the accretion disks cause outbursts (for reviews of
dwarf novae, see \cite{osa96review}).  SU UMa-type dwarf novae are a
class of DNe, which show superhumps during their long, bright outbursts
(superoutbursts) (for basic reviews, see \cite{vog80suumastars};
\cite{war85suuma}; \cite{war95suuma}).

   SU UMa-type dwarf novae have short orbital periods ($P_{\rm orb}$
usually shorter than 0.1 d), qualifying them as the highly evolved
population on the standard evolutionary track of CVs
(\cite{rap82CVevolution}; \cite{rap83CVevolution};
\cite{kin88binaryevolution}; \cite{kol99CVperiodminimum};
\cite{kin00CVevolution}).  Some objects have been even suggested to
have passed the CV minimum period, at which the thermal time scales
of the secondary star (Kelvin-Helmholtz time) become comparable to
the mass-transfer time scales
(\cite{pac71CVminimumperiod}; \cite{kin88binaryevolution}).
After passing this period minimum, the mass-losing secondary
stars are believed to become degenerate, and will become CVs with
brown-dwarf secondaries
(\cite{how97periodminimum}; \cite{pol98TOAD}; \cite{pat98evolution};
\cite{pat01SH}; \cite{pol02CVbrowndwarfproc}).  Several works have been
extensively made to observationally confirm possibility (\cite{cia98CVIR};
\cite{vantee99v592her}; \cite{how01llandeferi}; \cite{men01j1050};
\cite{ste01wzsgesecondary}; \cite{lit03CVBD}).  There seems to be
a certain degree of emerging evidence that at least some SU UMa-type dwarf
novae look like to have brown dwarf secondaries.

   These object are also intriguing objects from the standpoint of
the disk-instability model for dwarf nova outbursts.  Historically, the
most extreme member (WZ Sge) was selected as the prototype of a small
class of SU UMa-type dwarf novae (WZ Sge-type dwarf novae), which
are originally characterized by a long ($\sim$ 10 yr) outburst recurrence
time and a large ($\sim$ 8 mag) outburst amplitude (\cite{bai79wzsge};
\cite{dow81wzsge}; \cite{pat81wzsge}; \cite{odo91wzsge};
\cite{kat01hvvir}).  The origin of such powerful outbursts of WZ Sge
has still been one of the central problems of dwarf nova accretion
disks (\cite{sma93wzsge}; \cite{osa95wzsge}; \cite{las95wzsge};
\cite{war96wzsge}; \cite{min98wzsge}; \cite{mey98wzsge};
\cite{osa01egcnc}; \cite{bua02suumamodel}).

   From a slightly different standpoint, \citet{how95TOAD} observationally
proposed that some DNe with large outburst amplitudes (mostly SU UMa-type
dwarf novae) show unusual properties, and called them Tremendous Outburst
Amplitude Dwarf Novae (TOADs).  Although there exists an argument against
this nomenclature (cf. \cite{pat96alcom}), these objects are generally
considered to represent a borderline population of DNe between the most
unusual WZ Sge-type dwarf novae and usual SU UMa-type dwarf novae.
The TOADs and WZ Sge-type stars are also known to show extremely low
frequency of normal outbursts \citep{war95suuma}.  \citet{how95TOAD} and
\citet{how95swumabcumatvcrv} reported that the TOADs have unusual properties
of their superoutburst, particularly in that the TOADs sometimes show
{\it intermediate} outbursts having properties between full superoutbursts
and SU UMa-type normal outbursts, and in that some of the superoutbursts
of the TOADs are followed by post-superoutburst rebrightenings.
The latter property has been subsequently recognized as a common feature
with soft X-ray transients (black-hole transients) \citep{kuu96TOAD}.
The most dramatic manifestation of this phenomenon was seen in EG Cnc
which showed six successive post-superoutburst rebrightenings
(\cite{kat97egcnc}; \cite{mat98egcncqui}; \cite{pat98egcnc}).
\citet{how95TOAD} proposed that the unusual properties of the TOADs
are a result of low mass-transfer rate and low viscosity in quiescence.
This possibility has been tested by several authors to more precisely
reproduce the observed properties (\cite{osa95wzsge}; \cite{osa97egcnc};
\cite{mey98wzsge}; \cite{osa01egcnc}).

   From the observational side, both TOADs and WZ Sge-type dwarf novae
are known to (almost exclusively) show lengthening of the superhump period
during the superoutburst plateau (\cite{sem97swuma}; \cite{nog98swuma};
\cite{kat98super}; \cite{bab00v1028cyg}; \cite{kat01wxcet};
\cite{kat01hvvir}), whose origin is still poorly understood.
\citet{bab00v1028cyg} showed that a
significant deviation from the linear declining trend and a regrowth
of the superhumps during the terminal stage of a superoutburst in
V1028 Cyg, whose outburst amplitude is comparable to that of TOADs.
These phenomena, although still poorly described or understood,
may provide an additional clue for understanding the unusual behavior of
the accretion disks in large-amplitude, rarely outbursting SU UMa-type
dwarf novae, which are sometimes referred to as TOADs or WZ Sge-type.

   HO Delphini (= S 10066) is a dwarf nova discovered by \citet{hof67an29043}.
\citet{hof67an29043} reported two outburst in 1963 October and
1966 September.  Although the object was given the permanent variable star
designation \citep{NameList56}, virtually no observation had been reported
until recent years.  This lack of observation was partly because of the
\timeform{1D} error in the original discovery report by \citet{hof67an29043}.
The error was corrected in the third volume of the fourth edition of
the GCVS \citep{GCVS}.
\citet{DownesCVatlas1} gave a finding chart and \citet{kat99var5} reported
precise coordinates.  \citet{mun98CVspec5} recorded relatively strong
Balmer and He~{\sc i} emission lines, confirming the CV classification.
The He~{\sc ii} 4686\AA\, emission line was probably weakly present.

   The object has been regularly monitored by the VSNET\footnote{
$\langle$http://www.kusastro.kyoto-u.ac.jp/vsnet/$\rangle$} members
since 1994.  In spite of the monitoring, only three confirmed outbursts
have been recorded (1994 August--September, 1996 August--September,
2001 August--September).  The relatively low occurrence of the outbursts
is consistent with the initial finding by \citet{hof67an29043}.

   We observed HO Del during the two outbursts in 1994 and 2001.  We
also performed several snapshot observations during the 1996 outburst.

\section{Observation}

   The 1994 and 1996 observations were performed using a CCD camera
(Thomson TH~7882, 576 $\times$ 384 pixels, on-chip 2 $\times$ 2 binning
adopted) attached to the Cassegrain focus of the 60-cm reflector
(focal length = 4.8 m) at Ouda Station, Kyoto University \citep{Ouda}.
An interference filter was used which had been designed to reproduce the
Johnson $V$ band.  The frames were first corrected for standard de-biasing
and flat fielding, and were then processed by a microcomputer-based
photometry package developed by one of the authors (TK).
The 2001 observations were performed at
Nyr\"{o}l\"{a} Observatory using a 40-cm Schmidt--Cassegrain telescope and
an unfiltered ST-7E CCD camera.  The images were analyzed with the AIP4WIN
aperture photometry package.  The Ouda observations used GSC 1100.64
(GSC $V$ magnitude 12.76) as the comparison star, and GSC 1100.213
($V$ = 13.28) and GSC 1100.98 ($V$ = 13.76) as the check stars.
The Nyr\"{o}l\"{a} observations used GSC 1100.324 ($V$ = 12.16) as the
primary comparison star and GSC 1100.213 as the check star.
The constancy of the comparison stars during the runs was confirmed by
comparisons with the check stars.

   Barycentric corrections to the observed times were applied before the
following analysis.  The log of observations is summarized in table
\ref{tab:log}.

\begin{table*}
\caption{Journal of CCD photometry.}\label{tab:log}
\begin{center}
\begin{tabular}{ccrcccrccc}
\hline\hline
\multicolumn{3}{c}{Date}& Start--End$^*$ & Filter & Exp(s) & $N$
        & Mean mag$^\dagger$ & Error & Obs$^\ddagger$ \\
\hline
1994 & August & 26 & 49591.015--49591.119 & $V$ & 60 & 120 & 1.146 &
     0.005 & O \\
     &   & 27 & 49592.014--49592.219 & $V$ &  60 & 141 & 1.224 & 0.006 & O \\
     &   & 28 & 49593.065--49593.174 & $V$ &  90 &  91 & 1.412 & 0.003 & O \\
     &   & 29 & 49594.021--49594.233 & $V$ &  90 & 122 & 1.529 & 0.003 & O \\
     &   & 30 & 49595.006--49595.217 & $V$ & 120 &  59 & 1.607 & 0.010 & O \\
     &   & 31 & 49596.184--49596.218 & $V$ & 120 &  20 & 1.565 & 0.020 & O \\
     & September & 1 & 49597.024--49597.240 & $V$ & 120 &  99 & 1.674 &
     0.004 & O \\
     &   &  2 & 49598.137--49598.170 & $V$ & 120 &   7 & 1.843 & 0.084 & O \\
1996 & September & 8 & 50334.972--50334.976 & $V$ & 90 &  5 & 2.132 &
     0.032 & O \\
     &   & 15 & 50341.904--50341.905 & $V$ &  30 &   3 & 3.550 & 0.585 & O \\
2001 & August & 28 & 52150.296--52150.461 & none & 13 & 706 & 2.104 &
     0.004 & N \\
\hline
 \multicolumn{10}{l}{$^*$ BJD$-$2400000.} \\
 \multicolumn{10}{l}{$^\dagger$ Differential magnitudes to the comparison star.} \\
 \multicolumn{10}{l}{$^\ddagger$ O (Ouda), N (Nyr\"{o}l\"{a})} \\
\end{tabular}
\end{center}
\end{table*}

\section{Results}

\subsection{Course of Outburst}

\begin{figure}
  \begin{center}
%    \FigureFile(88mm,60mm){out2.eps}
    \FigureFile(88mm,60mm){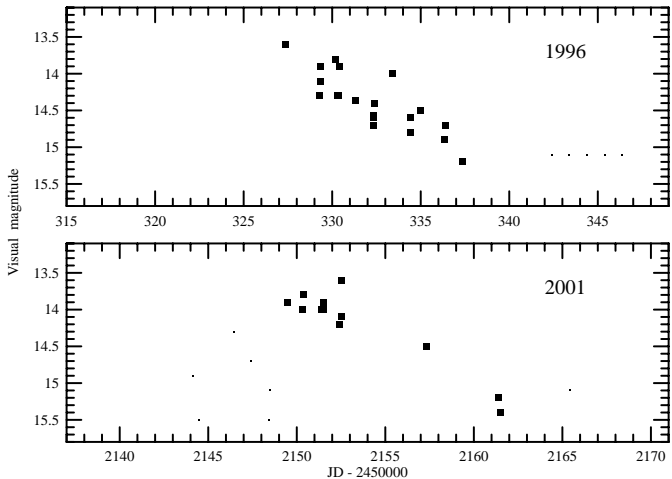}
  \end{center}
  \caption{Light curve of the 1996 August--September and
  2001 August--September superoutbursts.
  The large and small dots represent visual positive observations and
  upper limit observations, respectively, reported to VSNET.}
  \label{fig:out2}
\end{figure}

   The early stage of the 2001 August--September outburst (as shown in
subsection \ref{sec:sh}, both 1994 August--September outburst and 2001
August--September outburst were superoutbursts) was relatively well
followed by the VSNET observers.  The two superoutbursts in 1996 and 2001
are shown in figure \ref{fig:out2}.  Although the start and the termination
of each superoutbursts were not severely constrained, the apparent duration
($\sim$13 d) is characteristic to that of an ordinary SU UMa-type
superoutburst.

\begin{figure}
  \begin{center}
%    \FigureFile(88mm,60mm){out.eps}
    \FigureFile(88mm,60mm){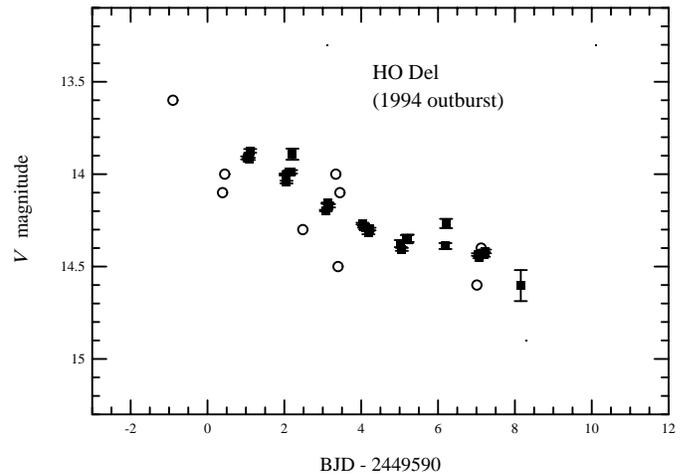}
  \end{center}
  \caption{Light curve of the 1994 August--September superoutburst.
  The filled squares with error bars are our $V$-band CCD observations.
  The open circles and small dots are visual observations and upper limit
  observations, respectively, reported to VSNET.}
  \label{fig:out}
\end{figure}

   Figure \ref{fig:out} shows the light curve of the 1994 August--September
superoutburst.  The duration of the outburst was longer than $\sim$10 d.
Based on the $V$-band CCD observations, the object linearly faded at
a rate of 0.14 mag d$^{-1}$ between BJD 2449591 and 2449594.  This rate
is quite characteristic to an SU UMa-type superoutburst plateau
(cf. \cite{war85suuma}; \cite{pat93vyaqr}; \cite{kat02v359cen}).
The rate of decline became smaller toward the late stage of the
superoutburst.  The rate reached a minimum of 0.03 mag d$^{-1}$ between
BJD 2449595 and 2449597.  Such a phenomenon was observed in V1028 Cyg
\citep{bab00v1028cyg}.

\subsection{Superhump Period}\label{sec:sh}

\begin{figure}
  \begin{center}
%    \FigureFile(88mm,60mm){ho28.eps}
    \FigureFile(88mm,60mm){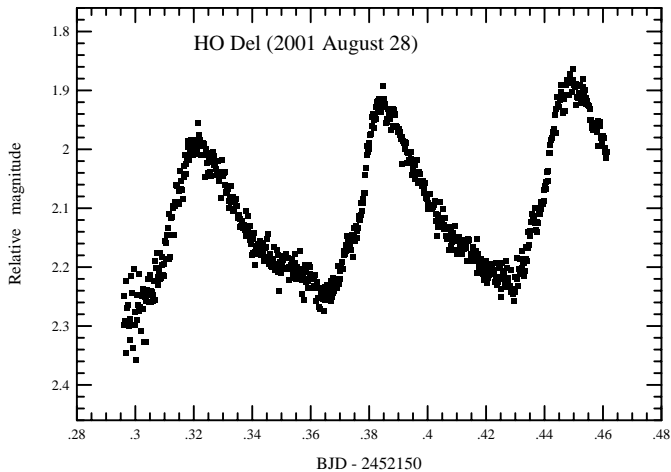}
  \end{center}
  \caption{Superhumps in HO Del observed on 2001 August 28.}
  \label{fig:ho28}
\end{figure}

   Superhumps were clearly present both in 1994 and 2001 observations,
clarifying the SU UMa-type nature of HO Del.
Figure \ref{fig:ho28} shows the best representative light curve of the
superhumps observed on 2001 August 28.  Since the 1994 observations had
shorter nightly coverages, we first determined an approximate superhump
period ($P_{\rm SH}$) from this longest and highest quality run, and
refine the period using the 1994 observation.  A period analysis of
the 2001 August 28 with the Phase Dispersion Minimization (PDM: \cite{PDM}),
after removing the linear trend, yielded a period of 0.0642(2) d.
The error of the period was estimated using the Lafler--Kinman class of
methods, as applied by \citet{fer89error}.

\begin{figure}
  \begin{center}
%    \FigureFile(88mm,60mm){pdm.eps}
    \FigureFile(88mm,60mm){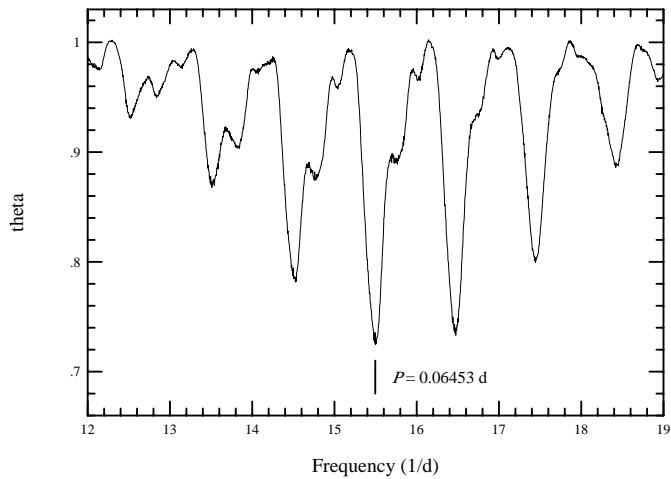}
  \end{center}
  \caption{Phase Dispersion Minimization analysis of the 2001 observation
  (August 26--29).  See text for the selection of the period.}
  \label{fig:pdm}
\end{figure}

   The well-observed portion of the 1994 observation (August 26--29)
were analyzed in a similar way, after removing the linear decline trend
of the outburst, and corrected for a small nightly deviations from the
linear trend.  The resultant theta diagram is shown in figure
\ref{fig:pdm}.  Among the possible one-day aliases, the frequency
15.496$\pm$0.013 d$^{-1}$, corresponding to $P_{\rm SH}$ = 0.06453(6) d,
is the only acceptable period in comparison with the 2001 observation.
This superhump period supersedes the previously reported preliminary
value cited in \citet{nog97sxlmi}.

\section{Superhump Profile}

\begin{figure}
  \begin{center}
%    \FigureFile(88mm,110mm){phase.eps}
    \FigureFile(88mm,110mm){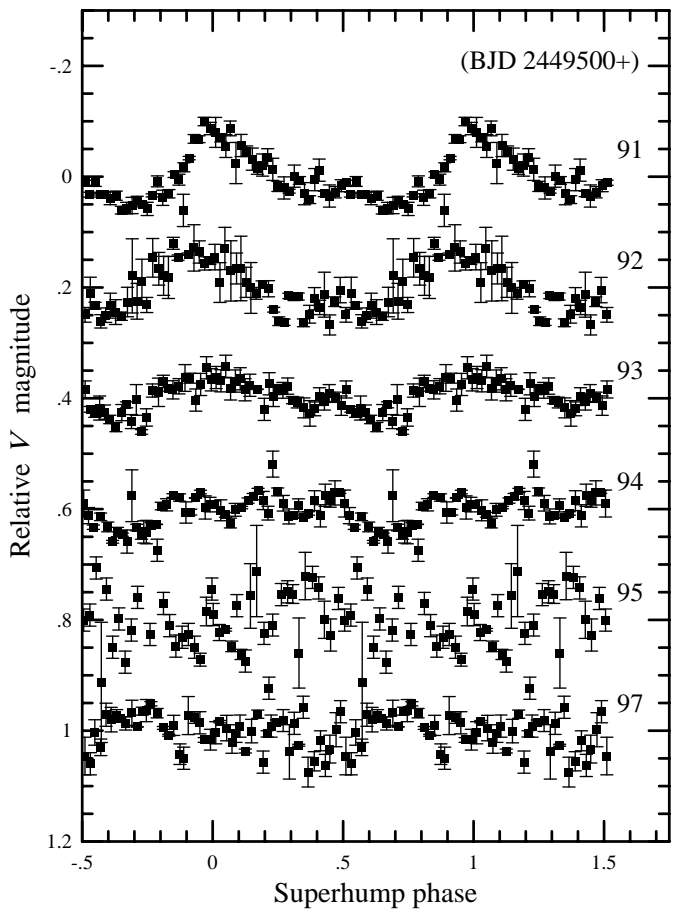}
  \end{center}
  \caption{Nightly averaged hump profiles during the 1994 outburst, folded
  by $P_{\rm SH}$ = 0.06453 d.  The phase zero corresponds to
  BJD 2449591.050.  The superhump signal decayed rather rapidly.  After
  BJD 2449594, the superhump signal almost disappeared.
  }
  \label{fig:phase}
\end{figure}

   The superhump profile on 2001 August 28 (the observation was performed
within 2 d of the outburst rise)\footnote{
  Initial outburst detection was made by M. Reszelski on August 27.972 UT at
  a visual magnitude of 13.9.  The object was not detected in outburst 1 d
  before this observation.  See vsnet-alert 6477,
  $\langle$http://www.kusastro.kyoto-u.ac.jp/vsnet/Mail/alert6000/msg00477.html$\rangle$.
}, was quite characteristic of fully developed SU UMa-type superhumps
(\cite{vog80suumastars}; \cite{war85suuma}).  The profiles were, however,
different during the 1994 outburst (figure \ref{fig:phase}).\footnote{
  The start of the 1994 August outburst was not well-constrained.  The
  initial outburst detection was made on 1994 August 24.600 UT at a visual
  magnitude of 13.6 by M. Moriyama.  No observations were reported for
  a month preceding this detection.  The reported magnitude, however,
  suggests that the outburst was detected during its relatively early
  stage.
}  The superhump signal decayed rather rapidly.  After BJD 2449594 (1994
August 29, $\sim$5 d after the initial outburst detection), the superhump
signal almost disappeared.

\section{Discussion}

\subsection{General Properties of Outbursts}\label{sec:out}

   In spite of the monitoring by the VSNET members, no confirmed normal
outbursts have been recorded between 1994 and 2002.  Since the maximum
of the superoutbursts reached a magnitude of 13.6, at least some of normal
outbursts, which are expected to have maximum magnitude of 14.1--14.6
\citep{war85suuma}, should have been recorded by modern instruments.
The lack of such detections seems to suggest that normal outbursts are
indeed rare in HO Del.  The shortest intervals of the recorded
superoutbursts was $\sim$740 d.  This interval suggests that superoutbursts
are also relatively rare, although some of the outbursts or superoutbursts
should have been missed during the unavoidable seasonal gaps.  The
small number of the recorded outbursts (likely superoutbursts)
inferred from the literature
\citep{hof67an29043} also seems to suggest a low outburst frequency.

   Among well-observed SU UMa-type dwarf novae, a small number of object
have comparable intervals of superoutbursts: UV Per ($\sim$960 d),
VY Aqr (350--800 d), WX Cet ($\sim$1000 d), SW UMa (459--954 d),
V1028 Cyg (380--450 d), V1251 Cyg ($\sim$1160 d),
EF Peg (1000--1400 d, VSNET; \cite{kat01hvvir}),
BC UMa ($>$1800 d or $\sim$1040 d, \cite{kat01hvvir},\footnote{See also
$\langle$http://www.kusastro.kyoto-u.ac.jp/vsnet/DNe/\\bcuma0302.html$\rangle$.
}
DV UMa ($\sim$770 d, \cite{nog01dvuma}),
V725 Aql ($\geq$1000 d, \cite{uem01v725aql})
Most of these objects are large-amplitude dwarf novae, sometimes classified
as TOADs (\cite{how95TOAD}).

   Other less well-documented systems with low occurrence
of normal outbursts and intervals of superoutbursts likely comparable
to HO Del include:
PU Per (\cite{kat95puper}; \cite{kat99puper}),
V844 Her (\cite{kat00v844her}; \cite{tho02gwlibv844herdiuma}),
QY Per (\cite{bus79VS17}; \cite{kat00qyperiauc}),
V359 Cen (\cite{kat02v359cen}),
and KV Dra (\cite{nog00kvdra}).

   The outburst cycle lengths and the apparently
low occurrence of normal outbursts in HO Del seem to share common
properties with the so-called TOADs or WZ Sge-type stars, although the
outburst amplitude ($\sim$5.0 mag) is smaller than those of the TOADs or
WZ Sge-type stars.  If the low outburst occurrence of HO Del is confirmed
by future more intensive observations, HO Del may be a rare object
with a long outburst cycle length and a rather normal outburst amplitude.

   A normal outburst amplitude naturally suggests a normal quiescent
luminosity (cf. \cite{war87CVabsmag}), which is indicative of a normal
mass-transfer rate.  In contrast, a long recurrence time would suggest
a low mass-transfer rate \citep{ich94cycle}.  If the mass-transfer rate
indeed turns out to be normal, the occurrence of outbursts may be somehow
suppressed in HO Del.

\subsection{Distance, proper motion, and X-ray Luminosity}

   By applying Warner's relation \citep{war87CVabsmag} using the newly
established $P_{\rm SH}$,\footnote{
  Since the difference between $P_{\rm SH}$ and the orbital period
  ($P_{\rm orb}$) is expected to be a few \% at this period
  \citep{StolzSchoembs}, we can safely use $P_{\rm SH}$ as a substitute
  for $P_{\rm SH}$ in applying to Warner's relation.
  \citet{pat98evolution} listed a likely orbital period without details.
  Although this period seems to be consistent with $P_{\rm SH}$, we probably
  need accurate $P_{\rm orb}$ determination by radial-velocity studies
  before making a final conclusion on the difference between $P_{\rm SH}$
  and $P_{\rm orb}$.
} the maximum $M_V$ is expected to be +5.2 (see also a discussion in
\cite{kat02v592her}).  An inclination effect \citep{war86NLabsmag} is
likely to be neglected because this star has a low to moderate inclination,
as inferred from the lack of eclipses and the single-peaked appearance of
the emission lines \citep{mun98CVspec5}.  Since Warner's relation
is expected to apply to normal outbursts of SU UMa-type dwarf novae
(cf. \cite{kat02v592her}; \cite{can98DNabsmag}), we use a $-$0.5 mag
correction to derive an expected maximum $M_V$ for superoutbursts.
Using the maximum visual magnitude of 13.6 for the recent superoutbursts,
the distance is estimated to be $\sim$600 pc.  Given the above
uncertainties, the error of the distance modulus is expected to be 0.3 mag,
which makes a range of the likely distance to be 400--800 pc.

   In order to study a possible proper motion, we examined the
available astrometric catalogs and the scanned images (Table
\ref{tab:astrometry}).  The errors are typically less than
\timeform{0.3''}.
Although the literal values of declination were slightly different,
no distinct proper motion was detected in a comparison of DSS 1 and 2 scans.
This indicates that the proper motion of HO Del is smaller than
\timeform{0.02''} yr$^{-1}$.  USNO B1.0 \citep{USNOB10} shows zero
proper motion for this object.
This lack of a detectable proper motion is consistent with the above
distance estimate.

\begin{table}
\caption{Astrometry of HO Del.}\label{tab:astrometry}
\begin{center}
\vspace{10pt}
\begin{tabular}{cccc}
\hline\hline
Source    & R. A. & Decl.               & Epoch \\
          & \multicolumn{2}{c}{(J2000.0)} & \\
\hline
USNO A2.0 & 20 36 55.514 & +14 03 10.15 & 1953.682 \\
USNO B1.0 & 20 36 55.485 & +14 03 09.74 & 1976.0   \\
GSC 2.2.1 & 20 36 55.498 & +14 03 09.46 & 1990.628 \\
\hline
\end{tabular}
\end{center}
\end{table}

   HO Del is quite reasonably identified with a ROSAT source
1RXS J203654.3+140301 \citep{ROSATFSCiauc}, which has a 52--201 keV count rate
of 0.030 count s$^{-1}$.  By following the formulation by \citep{ver97ROSAT},
we obtain an X-ray luminosity of 10$^{31.1\pm0.2}$ erg s$^{-1}$ (the error
includes both errors of the count rate and the distance estimate).
Although we will need future different ways of, or an even direct, distance
estimates, this value would potentially make HO Del one of the luminous
X-ray sources among dwarf novae \citep{ver97ROSAT}.  This
luminosity would suggest that the white dwarf of HO Del may be weakly
magnetic, as in intermediate polars (IPs) \citep{ver97ROSAT}.  If the inner
disk is truncated by the magnetized white dwarf in an IP, the apparently
low occurrence of outbursts in HO Del (subsection \ref{sec:out}) may be
a result of resulting suppression of the disk instability
\citep{ang89DNoutburstmagnetic}.  Although the strength of the quiescent
He~{\sc ii} emission is not as striking as in other typical intermediate
polars, a search for IP-type coherent oscillations would be meaningful.

\section{Statistics and Late-Time Superhump Evolution in SU UMa-Type
         Dwarf Novae}

   HO Del showed a relatively rapid decay of the superhump amplitudes.
A similar phenomenon was observed in V1028 Cyg \citep{bab00v1028cyg} and
DM Dra \citep{kat02dmdra}.  The subsequent behavior was different between
HO Del and V1028 Cyg in that no regrowth of the superhumps during the
later stage, during which the rate of decline became temporarily slower,
was observed in contrast to the V1028 Cyg case \citep{bab00v1028cyg}.

\begin{figure}
  \begin{center}
%    \FigureFile(88mm,120mm){reb.eps}
    \FigureFile(88mm,120mm){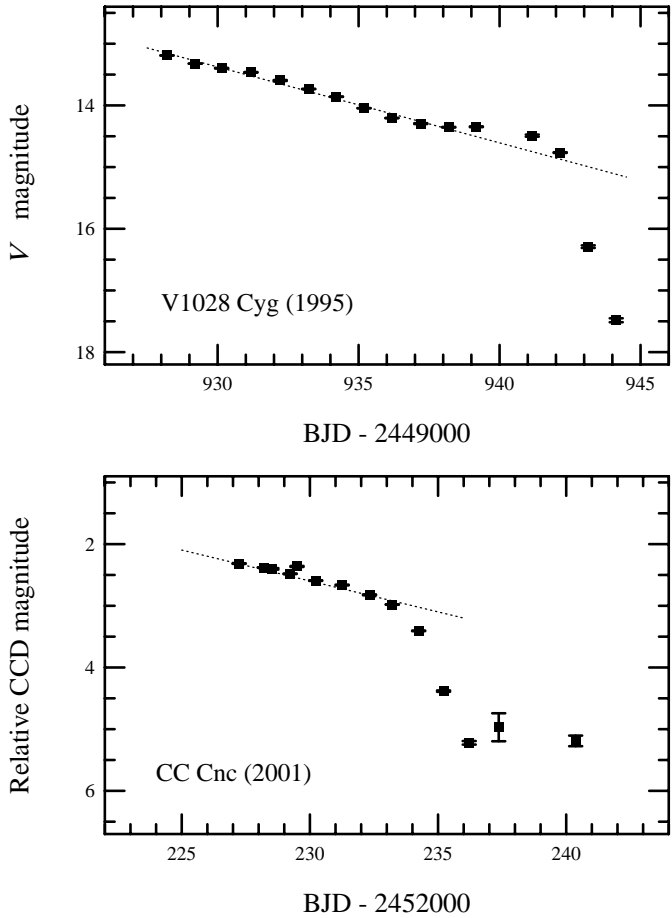}
  \end{center}
  \caption{Brightening trend near the termination of a superoutburst.
  [Upper: with brightening (V1028 Cyg: data from \cite{bab00v1028cyg}).
  Lower: without brightening (CC Cnc: data from \citet{kat02cccnc},
  the first night data were omitted because of a different instrument
  emplyed)].  The dashed lines represent linear fits to the earlier
  stage of the plateau portion.  The upward deviation in V1028 Cyg is
  evident.}
  \label{fig:reb}
\end{figure}

\begin{figure}
  \begin{center}
%    \FigureFile(75mm,150mm){v1028sh.eps}
    \FigureFile(75mm,150mm){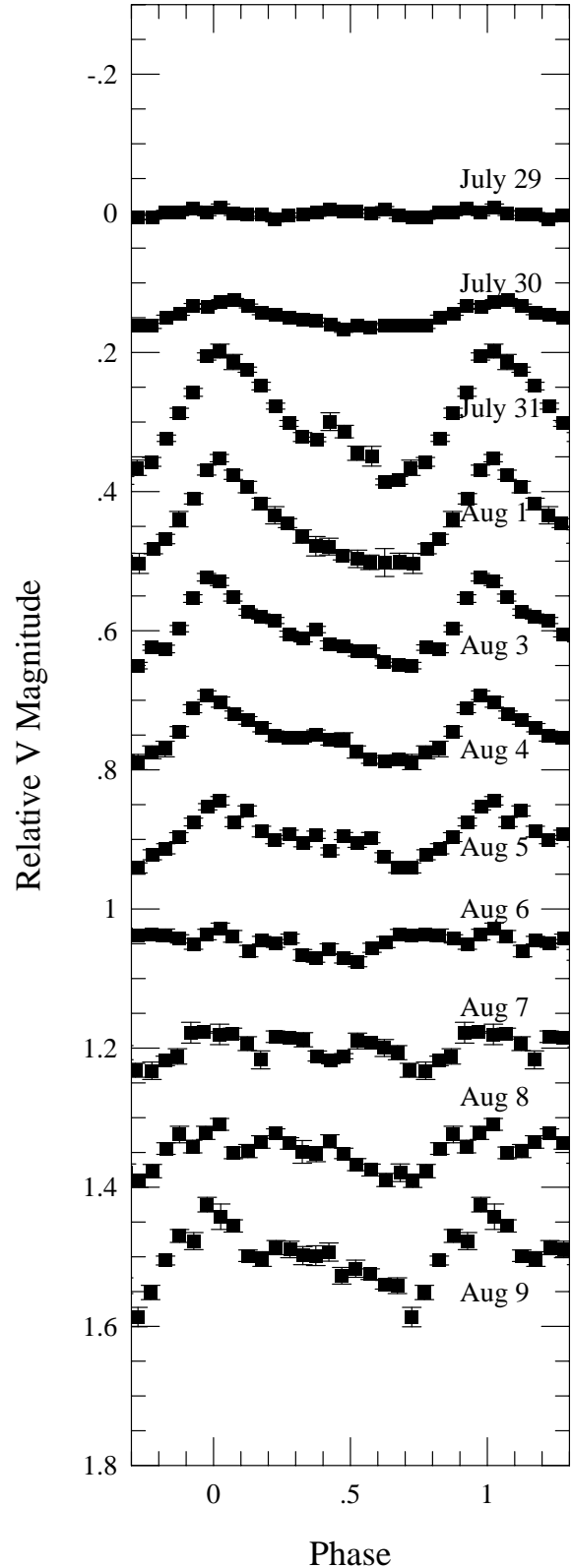}
  \end{center}
  \caption{Example of regrowth of superhumps near the termination of
  a superoutburst [reproduced from Figure 6 in \citet{bab00v1028cyg}
  by permission].}
  \label{fig:v1028sh}
\end{figure}

   Since it has been becoming evident that the regrowth of the superhumps
is rather commonly seen in SU UMa-type dwarf novae with infrequent
outbursts, we performed a literature survey of SU UMa stars in order
to see the relation between this phenomenon, as well as the brightening
trend near the termination of a superoutburst, and the other parameters.
Figure \ref{fig:reb} shows a comparison of systems with and without
brightening trend near the termination of a superoutburst.
\citet{ish03hvvir} also provides a well-recorded presentation of the
same phenomenon in HV Vir.
Figure \ref{fig:v1028sh} shows a representative example of regrowth of
superhumps near the termination of a superoutburst \citep{bab00v1028cyg}.

   The result is presented in table \ref{tab:regrowth}.  The ``Yes''
classification in the ``Rise'' field (brightening near the termination of
a superoutburst) generally corresponds to more than 0.1 mag deviation from
the linear decline trend.  The ``Yes'' classification in the ``Regrowth''
field (regrowth of the superhumps during the terminal stage of
a superoutburst) corresponds to a detectable regrowth of the superhumps
{\it before} the start of the final, rapid decline.  A question mark (?)
follows if observations did not sufficiently covered the late stages
of a superoutburst.
Since it is not well-known whether the same star exhibits the same
evolution of late-time outburst behavior and superhumps during different
superoutbursts, we tried to examine each superoutburst separatedly
whenever possible.  In some cases, a combined result
from a set of less favorably covered different superoutburst is presented.
In the table and the following discussion, we, in principle, used
$P_{\rm SH}$ values, which were better determined than $P_{\rm orb}$ values
in many systems.

   In examining correlations between $P_{\rm SH}$ and other parameters,
we predicted the $P_{\rm SH}$'s of these objects based on the empirical
relation \citep{mol92SHexcess}.

   In ER UMa stars (ER UMa, V1159 Ori, RZ LMi, DI UMa and IX Dra: see
\cite{kat95eruma}; \cite{nog95rzlmi}; \cite{kat99erumareview}), we
did not examine the evolution of the superhumps of these objects
in the same manner, because there is emerging evidence of unusual
superhump evolution at least in one of these systems \citep{kat03erumaSH},
suggesting that superhumps in ER UMa (and possibly in other ER UMa stars)
are different in nature from other SU UMa-type dwarf novae.

   So-called helium dwarf novae
(CR Boo: \cite{woo87crboo}; \cite{pat97crboo}; \cite{pro97crboo};
\cite{kat00crboo}; \cite{kat01crboo},
V803 Cen: \cite{odo87v803cen}; \cite{odo89v803cen}; \cite{pat00v803cen};
\cite{kat00v803cen}; \cite{kat01v803cen},
CP Eri: \cite{abb92alcomcperi}; \cite{gro01cperi},
KL Dra: \cite{woo02kldra}) were not included in this survey, although
two unusually short-period, hydrogen rich systems (V485 Cen and
EI Psc = 1RXS J232953.9+062814) were included.

   The typical superhump period and supercycle length ($T_s$) are given in
the initial line of each object.  These data are usually taken from
\citet{nog97sxlmi}, supplemented with new measurements based on the
references listed.

\begin{table*}
\caption{Outburst parameters and superhump regrowth in SU UMa-type stars.}
\label{tab:regrowth}
\begin{center}
\begin{tabular}{llcccc}
\hline\hline
Object$^*$ & $P_{\rm SH}$ & $T_s$$^\dagger$ & Rise$^\ddagger$ &
       Regrowth$^\S$ & References \\
\hline
V485 Cen (1997) & 0.04216 & 320: & No & No & 1 \\
EI Psc (2001)   & 0.04637 & $\cdots$ & No & No &
                                             2,
                                             3,
                                             4 \\
DI UMa (1996)   & 0.05529 & 25--45 & No & ER & 5 \\
\phantom{DI UMa} (1998) & &        & No & $\cdots$ &
                                             6 \\
V844 Her (2000+) & 0.05592 & 220--290 & Yes? & $\cdots$ &
                                             7,
                                             8,
                                             9 \\
V2176 Cyg (1996) & 0.0561 & $>$2000 & No$^\|$ & No? &
                                             10 \\
V592 Her (1998) & 0.056498 & 4000: & Yes? & Yes? &
                                             11,
                                             12 \\
\phantom{V592 Her} (1968) & &      & Yes? & $\cdots$ &
                                             13,
                                             14 \\
UW Tri (1995)   & 0.0569   & 4000: & $\cdots$ & $\cdots$ &
                                             15 \\
LL And (1993)   & 0.057006 & 5000 & No? & $\cdots$ &
                                             16,
                                             17 \\
WZ Sge (2001)   & 0.05721 & 8200--11000 & No$^\|$ & No &
                                             18,
                                             19,
                                             20 \\
AL Com (1995)   & 0.05735 & 7000  & No$^\|$ & No? &
                                             21,
                                             22,
                                             23,
                                             24 \\
MM Hya (2001)   & 0.05785 & 380   & Yes  & $\cdots$ &
                                             25 \\
PU CMa (2003+)  & 0.05789  & 362--391 & No & No? &
                                             26,
                                             27,
                                             28 \\
HV Vir (1992)   & 0.05820 & 3800  & Yes? & $\cdots$ &
                                             29,
                                             30 \\
\phantom{HV Vir} (2001) & &       & Yes  & Yes  &
                                             31 \\
SW UMa (1996)   & 0.05833 & 459--954 & Yes? & Yes? &
                                             32,
                                             33,
                                             34 \\
\phantom{SW UMa} (2002) & &          & Yes  & Yes  &
                                             35 \\
V1141 Aql (2002) & 0.05930 & 340:    & $\cdots$ & $\cdots$ &
                                             36,
                                             37 \\
RZ LMi (1995+)  & 0.05946 &   19     & No   & ER  &
                                             38,
                                             39 \\
WX Cet (1998)   & 0.05949 & 1000:    & Yes  & Yes  & 40 \\
\phantom{WX Cet} (1991) & &          & Yes? & Yes? & 41 \\
\phantom{WX Cet} (1989) & &          & Yes? & $\cdots$ &
                                             42 \\
KV Dra (2002)   & 0.06012 & 840:    & Yes & Yes? &
                                             43,
                                             44 \\
\phantom{KV Dra} (2000) &  &         & Yes? & $\cdots$ &
                                             45,
                                             46 \\
T Leo  (1993)   & 0.0602  &  420     & $\cdots$ & $\cdots$ &
                                             47,
                                             48,
                                             49 \\
EG Cnc (1996)   & 0.06038 & 7000     & No?  & No?  &
                                             50,
                                             51,
                                             52 \\
RX Vol (2003)   & 0.06117 & $\cdots$ & Yes   & Yes &
                                             53 \\
MM Sco (2002)   & 0.06136 & 298--497 & $\cdots$ & $\cdots$ &
                                             54 \\
V1028 Cyg (1995) & 0.06154 & 380--450 & Yes  & Yes  &
                                             55 \\
UZ Boo (1994)   & 0.0619: & 5800?    & No? & $\cdots$ &
                                             30,
                                             56,
                                             57,
                                             58 \\
V1040 Cen (2002) & 0.0622 & 211 & No? & $\cdots$ &
                                             59,
                                             60 \\
AQ Eri (1991)   & 0.06225 & 300:     & $\cdots$ & $\cdots$ &
                                             61 \\
CI UMa (1995,2003) & 0.06264  & 140  & No? & $\cdots$ &
                                             62,
                                             63 \\
XZ Eri (2003)   & 0.06281 & 396:     & Yes & Yes &
                                             64,
                                             65 \\
GO Com (2003)   & 0.06297 & (2900:/$N$?) & Yes? & No? &
                                             66 \\
V402 And (2000,1985) & 0.06339 & $\cdots$ & Yes? & $\cdots$ &
                                             67,
                                             68,
                                             69 \\
CG CMa (1999)   & 0.0636: & $\cdots$ & Yes?$^\|$ & $\cdots$ &
                                             70,
                                             71 \\
V436 Cen (1978) & 0.06383 & 630      & Yes? & $\cdots$ &
                                             72,
                                             73 \\
V1159 Ori (1995+) & 0.0641 & 44.6--53.3 & No? & ER &
                                             39,
                                             74,
                                             75,
                                             76,
                                             77 \\
V2051 Oph (1998) & 0.06423 & 227     & No?  & No? &
                                             78,
                                             79 \\
VY Aqr (1986)   & 0.0645  & 350--800? & Yes? & Yes? &
                                             80 \\
BC UMa (2000)   & 0.06452 & $>$1800  & $\cdots$ & $\cdots$ &
                                             81 \\
HO Del (1994+)  & 0.06453 & 740:     & Yes & No &
                                             82 \\
OY Car (1980)   & 0.064544 & 346     & Yes? & Yes? &
                                             83,
                                             84 \\
\phantom{OY Car} (1985) &  &         & Yes? & $\cdots$ &
                                             85 \\
EK TrA (1985)   & 0.0649  &  485     & Yes & $\cdots$ &
                                             86 \\
1RXP J113123+4322.5 (2002) & 0.06495 & 370 & Yes? & $\cdots$ &
                                             87,
                                             88 \\
TV Crv (1994)   & 0.0650  & 390--450 & Yes? & Yes? & 89 \\
\phantom{TV Crv} (2001) & &          & Yes & Yes & 90 \\
\hline
 \multicolumn{6}{l}{$^*$ Year of the outburst in parentheses.} \\
 \multicolumn{6}{l}{$^\dagger$ Typical supercycle length (d), mainly taken from
\citet{nog97sxlmi}.} \\
 \multicolumn{6}{l}{\phantom{$^\dagger$} The others are estimates from VSNET
observations.} \\
 \multicolumn{6}{l}{$^\ddagger$ Brightening near the termination of superoutburst.} \\
 \multicolumn{6}{l}{$^\S$ Regrowth of the superhumps during the terminal stage of superoutburst.} \\
 \multicolumn{6}{l}{\phantom{$^\S$} ER: ER UMa stars, not examined (see text)} \\
 \multicolumn{6}{l}{$^\|$ Before the ``dip'' fading in WZ Sge-type dwarf novae.} \\
\end{tabular}
\end{center}
\end{table*}

\begin{table*}
\addtocounter{table}{-1}
\caption{(continued)}
\begin{center}
\begin{tabular}{llcccc}
\hline\hline
Object & $P_{\rm SH}$ & $T_s$ & Rise &
       Regrowth & References \\
\hline
ER UMa (1995+)  & 0.06556 &   43     & No? & ER &
                                             39,
                                             91,
                                             92 \\
AQ CMi (1996+)  & 0.06625 &  410?    & $\cdots$ & $\cdots$ &
                                             93,
                                             94,
                                             95 \\
UV Per (2001)   & 0.06641 &  960     & Yes? & Yes? & 96 \\
\phantom{UV Per} (1989) & &          & $\cdots$ & $\cdots$ &
                                             97 \\
CT Hya (1999)   & 0.06643 &  280     & No? & $\cdots$ &
                                             98 \\
\phantom{CT Hya} (1995) & &          & No? & No? & 99 \\
\phantom{CT Hya} (2000) & &          & No? & No? & 100 \\
\phantom{CT Hya} (2002) & &          & No? & No? & 100 \\
IX Dra (2001+)  & 0.06700 &  53      & No? & ER &
                                             101,
                                             102 \\
DM Lyr (1996)   & 0.0673  & 250      & No? & $\cdots$ &
                                             103 \\
\phantom{DM Lyr} (1998) & &          & No? & $\cdots$ &
                                             103 \\
AK Cnc (1995)   & 0.06749 & 210--390 & Yes? & Yes? &
                                             104 \\
\phantom{AK Cnc} (1992) & &          & $\cdots$ & $\cdots$ &
                                             105 \\
SS UMi (1998)   & 0.06778 & 84.7     & No & No &
                                             106,
                                             107,
                                             108 \\
SX LMi (1994)   & 0.06850 & 279      & Yes? & No? &
                                             109,
                                             110 \\
V701 Tau (1995) & 0.0689  & 380:     & $\cdots$ & $\cdots$ &
                                             112 \\
V550 Cyg (2000) & 0.0689: & $\cdots$ & $\cdots$ & $\cdots$ &
                                             113 \\
V1504 Cyg (2001) & 0.0690 & 137 & No & $\cdots$ &
                                             106,
                                             114,
                                             115 \\
KS UMa (1998,2003) & 0.07009 & 254   & No? & $\cdots$ &
                                             116,
                                             117,
                                             118 \\
RZ Sge (1996)   & 0.07042 & 266      & No  & No & 119 \\
\phantom{RZ Sge} (1994) & &          & No? & No? & 120 \\
\phantom{RZ Sge} (1981) & &          & No? & Yes? & 121 \\
V1208 Tau (2000) & 0.07060 & $\cdots$ & No? & No? &
                                             122 \\
IR Gem (1982)   & 0.07076 & 183      & $\cdots$ & $\cdots$ &
                                             106,
                                             123 \\
\phantom{IR Gem} (1991) & &          & $\cdots$ & $\cdots$ &
                                             124 \\
TY Psc (2001)   & 0.0708  & 210      & No? & $\cdots$ &
                                             125,
                                             126 \\
CY UMa (1999)   & 0.07210  & 300      & No? & No? & 127 \\
\phantom{CY UMa} (1995) &  &          & No? & No? & 128 \\
PU Per (1998)   & 0.0733  & $>$500   & $\cdots$ & $\cdots$ &
                                             129 \\
FO And (1994)   & 0.07411 & 100--140 & No? & No? &
                                             106,
                                             130 \\
VW CrB (1997+)  & 0.0743  & 240--370 & No? & $\cdots$ &
                                             131,
                                             132,
                                             133 \\
KV And (1994)   & 0.07434 & 240      & No? & No? & 134 \\
\phantom{KV And} (2002) & &          & No  & No? & 135 \\
AW Sge (2000)   & 0.0745  & $\cdots$ & $\cdots$ & $\cdots$ &
                                             136 \\
NSV 10934 (2003) & 0.07478 & $\cdots$ & No? & No? &
                                             54 \\
V368 Peg (1999,2000) & 0.075 & 390?  & No? & No? &
                                             137,
                                             138,
                                             139 \\
OU Vir (2003+)  & 0.07505  & 410:/$N$ & No? & No? &
                                              140,
                                              141,
                                              142 \\
CC Cnc (2001)   & 0.075518 & 370     & No & No & 143 \\
\phantom{CC Cnc} (1996) &  &         & No? & $\cdots$ &
                                                 144 \\
DM Dra (2000,2003) & 0.07567 & 440:  & No? & No? &
                                             145,
                                             146,
                                             147 \\
VZ Pyx (1996,1998) & 0.07568 & 270   & No? & $\cdots$ &
                                             148,
                                             149 \\
IY UMa (2000)   & 0.07588 & 285.5    & No  & No  &
                                             79,
                                             150,
                                             151,
                                             152 \\
\phantom{IY UMa} (2002) & &          & No  & No? &
                                             152 \\
AY Lyr (1987)   & 0.07597 & 210      & No  & No  &
                                             153,
                                             154 \\
\phantom{AY Lyr} (1977) & &          & No  & No  & 155 \\
V1251 Cyg (1991) & 0.07604 & 1160    & $\cdots$ & $\cdots$ &
                                             156 \\
HT Cas (1985)    & 0.076077 & $>$450? & $\cdots$ & $\cdots$ &
                                             157,
                                             158,
                                             159 \\
QY Per (1999)    & 0.07681 & $>$1800? & No  & No  &
                                             160 \\
QW Ser (2002)    & 0.07709 & 230--276 & Yes? & $\cdots$ &
                                             161,
                                             162,
                                             163 \\
\phantom{QW Ser} (2000) &  &         & No? & $\cdots$ &
                                             161 \\
BE Oct (1996)    & 0.07711 & $\cdots$ & $\cdots$ & $\cdots$ &
                                             164 \\
\hline
\end{tabular}
\end{center}
\end{table*}

\begin{table*}
\addtocounter{table}{-1}
\caption{(continued)}
\begin{center}
\begin{tabular}{llcccc}
\hline\hline
Object & $P_{\rm SH}$ & $T_s$ & Rise &
       Regrowth & References \\
\hline
VW Hyi (1972)    & 0.07714 &  179    & No  & No  & 165 \\
\phantom{VW Hyi} (1974) &  &         & No  & No  & 166 \\
\phantom{VW Hyi} (1978) &  &         & No  & No  & 167 \\
\phantom{VW Hyi} (1995) &  &         & No  & No  & 168 \\
WX Hyi (1977)    & 0.07737 &  174    & No  & No  &
                                              106,
                                              169 \\
Z Cha (1984+)     & 0.07740 &  287   & No  & No  &
                                              170,
                                              171,
                                              172,
                                              173 \\
\phantom{Z Cha} (1987) &   &         & No  & No & 174 \\
TT Boo (1993)     & 0.07811  & 245   & $\cdots$ & $\cdots$ &
                                              175 \\
WY Tri (2001)     & 0.078483 & $\cdots$ & $\cdots$ & $\cdots$ &
                                              176,
                                              177 \\
RZ Leo (2000)     & 0.078529 & 5800: & Yes & No? &
                                              178 \\
V630 Cyg (1996,1999) & 0.0789 & 145--$>$290 & No & $\cdots$ &
                                              179 \\
SU UMa (1989)     & 0.07904  & 160   & No  & No? & 180 \\
V1113 Cyg (1994)  & 0.0792   & 189.8 & No? & $\cdots$ &
                                              181,
                                              182 \\
SDSSp J173008.38+624754.7 (2001) & 0.07941 & 327: & No? & No? &
                                              183,
                                              184,
                                              185 \\
AW Gem (1995)     & 0.07943  & 300   & No? & $\cdots$ &
                                              186 \\
CU Vel            & 0.07999  & 700--900 & $\cdots$ & $\cdots$ &
                                              187,
                                              188 \\
DH Aql (2002)     & 0.08003  & 300   & No & No & 189 \\
\phantom{DH Aql} (1994) &    &       & $\cdots$ & $\cdots$ &
                                              190 \\
PV Per (1996)     & 0.0805   & 180?  & $\cdots$ & $\cdots$ &
                                              191,
                                              192 \\
HS Vir (1996)     & 0.08077  & 186 or 371 & No? & No? &
                                              193,
                                              194 \\
V359 Cen (2002+)  & 0.08092  & 307--397 & $\cdots$ & $\cdots$ &
                                              65,
                                              195 \\
V660 Her (1999,1995) & 0.081    & 380: & No & $\cdots$ &
                                              28,
                                              196,
                                              197,
                                              198 \\
V503 Cyg (1994,1998) & 0.08101  &  89 & No & No? &
                                              199,
                                              200,
                                              201 \\
BR Lup (1986--2003) & 0.08220 &  140 & $\cdots$ & No? &
                                              202,
                                              203,
                                              204 \\
NSV 09923 (2003)  & 0.08253  & $\cdots$ & $\cdots$ & $\cdots$ &
                                              205 \\
RX Cha (1998)     & 0.0839   & 430: & $\cdots$ & $\cdots$ &
                                              206,
                                              207 \\
V877 Ara (2002)   & 0.08411  & 285--374 & No? & No? &
                                              27 \\
AB Nor (2002)     & 0.08438   & 880:/$N$ & $\cdots$ & $\cdots$ &
                                              54 \\
CP Dra (2001+)    & 0.08474  & 400? & No? & $\cdots$ &
                                              208,
                                              209 \\
V369 Peg (1999)   & 0.08484  & 320: & No? & No? &
                                              68,
                                              210 \\
TU Crt (1998,2001+) & 0.0854  & 380--530 & No? & $\cdots$ &
                                              211,
                                              212,
                                              213,
                                              214 \\
HV Aur (1994,2002) & 0.08559  & $\cdots$ & $\cdots$ & $\cdots$ &
                                              215,
                                              216 \\
V589 Her (2002)   & 0.0864   & $\cdots$ & $\cdots$ & $\cdots$ &
                                              217 \\
EF Peg (1991)     & 0.08705  & 1000--1400 & No? & No? &
                                              218,
                                              219,
                                              220 \\
\phantom{EF Peg} (1997) &    &      & No? & No? &
                                              221 \\
TY PsA (1982,1984) & 0.08765  & 220 & No? & No? &
                                              222,
                                              223,
                                              224 \\
BF Ara (2002+)    & 0.08797  & 84.3 & No & No &
                                              225,
                                              226,
                                              227 \\
KK Tel (2002)     & 0.08803  &  394 & $\cdots$ & $\cdots$ &
                                              27 \\
V452 Cas (1999)   & 0.0881   & 320/$N$? & No & $\cdots$ &
                                              228,
                                              229 \\
DV UMa (1999)     & 0.08869  &  770 & Yes? & No? &
                                              30,
                                              230,
                                              231 \\
\phantom{DV UMa} (1997) &    &      & No? & $\cdots$ &
                                              232 \\
V419 Lyr (1999)   & 0.0901   & 340: & No? & No? &
                                              233,
                                              234 \\
UV Gem (2002)     & 0.0902   & 144? & $\cdots$ & $\cdots$ &
                                              235,
                                              236,
                                              237 \\
V344 Lyr (1993)   & 0.09145  & 109.6 & No? & No? &
                                              106,
                                              238 \\
YZ Cnc (1978,1988+) & 0.09204  & 134 & No  & No &
                                              155,
                                              239,
                                              240,
                                              241,
                                              242 \\
GX Cas (1994)     & 0.09297  &  307 & No & No &
                                              233,
                                              243 \\
\phantom{GX Cas} (1999) &    &      & No & No &
                                              244 \\
V725 Aql (1999)   & 0.09909  & $\geq$1000 & Yes? & Yes? &
                                              245,
                                              246 \\
\phantom{V725 Aql} (1995) &  &            & Yes? & Yes? &
                                              247 \\
MN Dra (2001--2003) & 0.1055   &   60  & No & No? &
                                              248,
                                              249 \\
NY Ser (1996+)    & 0.1064   & 70--100 & No & No &
                                              250 \\
TU Men (1980)     & 0.1262   & 600? & Yes? & No? &
                                              251,
                                              252,
                                              253,
                                              254 \\
\hline
\end{tabular}
\end{center}
\end{table*}

\begin{table*}
{\footnotesize
  {\bf References and remarks to table \ref{tab:regrowth}:}
  %% V485 Cen
    1. \citet{ole97v485cen};
  %% EI Psc
    2. \citet{uem02j2329letter};
    3. \citet{uem02j2329};
    4. \citet{tho02j2329};
  %% DI UMa
    5. \citet{kat96diuma};
    6. \citet{fri99diuma};
  %% V844 Her
    7. \citet{kat00v844her};
    8. \citet{ant96newvar};
    9. \citet{tho02gwlibv844herdiuma};
  %% V2176 Cyg
    10. \citet{nov01v2176cyg};
  %% V592 Her
    11. \citet{kat02v592her};
    12. \citet{due98v592her};
    13. \citet{ric68v592her};
    14. \citet{ric92wzsgedip};
  %% UW Tri
    15. \citet{kat01uwtri};
  %% LL And
    16. \citet{how96lland};
    17. T. Kato et al., in preparation;
  %% WZ Sge
    18. \citet{ish02wzsgeletter};
    19. \citet{pat02wzsge};
    20. R. Ishioka et al., in preparation;
  %% AL Com
    21. \citet{pyc95alcom};
    22. \citet{pat96alcom};
    23. \citet{nog97alcom};
    24. \citet{how96alcom};
  %% MM Hya
    25. R. Ishioka et al., in preparation and the VSNET observations
    (1997--2001);
  %% PU CMa
    26. T. Kato et al., vsnet-campaign-dn 3708 and VSNET observations;
    27. \citet{kat03v877arakktelpucma};
    28. \citet{tho03kxaqlftcampucmav660herdmlyr};
  %% HV Vir
    29. \citet{lei94hvvir};
    30. \citet{kat01hvvir};
    31. \citet{ish03hvvir};
  %% SW UMa
    32. \citet{sem97swuma};
    33. \citet{nog98swuma};
    34. \citet{how95swumabcumatvcrv};
    35. K. Tanabe et al., in preparation;
  %% V1141 Aql
    36. \citet{ole03v1141aql};
    37. VSNET observations;
  %% RZ LMi
    38. \citet{nog95rzlmi};
    39. \citet{rob95eruma};
  %% WX Cet
    40. \citet{kat01wxcet};
    41. \citet{kat95wxcet};
    42. \citet{odo91wzsge};
  %% KV Dra
    43. M. Uemura et al., vsnet-campaign-dn 2802, 2820;
    44. VSNET observations;
    45. \citet{nog00kvdra};
    46. \citet{van00kvdra};
  %% T Leo
    47. \citet{lem93tleo};
    48. \citet{kat97tleo};
    49. \citet{how99tleo};
  %% EG Cnc
    50. \citet{kat97egcnc};
    51. \citet{mat98egcncqui};
    52. \citet{pat98egcnc};
  %% RX Vol
    53. T. Kato et al., vsnet-campaign-dn 3637,3644;
  %% MM Sco
    54. \citet{kat03nsv10934mmscoabnorcal86};
  %% V1028 Cyg
    55. \citet{bab00v1028cyg};
  %% UZ Boo
     56. \citet{kuu96TOAD};
     57. \citet{ric92wzsgedip};
     58. \citet{bai79wzsge};
  %% V1040 Cen
    59. B. Monard, private communication, cf. vsnet-alert 7269;
    60. VSNET observations;
  %% AQ Eri
    61. \citet{kat91aqeri};
  %% CI UMa
    62. \citet{nog97ciuma};
    63. vsnet-campaign-dn 3588;
  %% XZ Eri
    64. \citet{uem03xzeri};
    65. \citet{wou01v359cenxzeriyytel};
  %% GO Com
    66. Superhump period variable (e.g. vsnet-campaign-dn 3768).
    Mean superhump period given;
  %% V402 And
    67. T. Kato et al., vsnet-campaign-dn 170;
    68. \citet{ant98v1008herv402andv369peg};
    69. VSNET observations;
  %% CG CMa
    70. \citet{kat99cgcma};
    71. \citet{due99cgcma};
  %% V436 Cen
    72. \citet{sem80v436cen};
    73. \citet{war75v436cen};
  %% V1159 Ori
    74. \citet{pat95v1159ori};
    75. \citet{nog95v1159ori};
    76. \citet{szk99v1159ori};
    77. \citet{kat01v1159ori};
  %% V2051 Oph
    78. \citet{kiy1998v2051oph};
    79. \citet{kat01v2051ophiyuma};
  %% VY Aqr
    80. \citet{pat93vyaqr};
  %% BC UMa
    81. VSNET Collaboration data;
  %% HO Del
    82. this paper;
  %% OY Car
    83. \citet{sch86oycar};
    84. \citet{krz85oycarsuper};
    85. \citet{nay87oycar};
  %% EK TrA
    86. \citet{has85ektra};
  %% J1131
    87. M. Uemura et al., vsnet-alert 7231;
    88. T. Kato and M. Uemura, vsnet-campaign-dn 3540;
  %% TV Crv
    89. \citet{how96tvcrv};
    90. R. Ishioka et al., in preparation;
  %% ER UMa
    91. \citet{kat95eruma};
    92. \citet{kat03erumaSH};
  %% AQ CMi
    93. J. Patterson, vsnet-alert 327;
    94. D. Nogami, vsnet-alert 328;
    95. VSNET observations;
  %% UV Per
    96. VSNET Collaboration data;
    97. \citet{uda92uvper};
  %% CT Hya
    98. \citet{kat99cthya};
    99. \citet{nog96cthya};
    100. D. Nogami et al., in preparation;
  %% IX Dra
    101. \citet{ish01ixdra};
    102. \citet{klo95exdraixdra};
  %% DM Lyr
    103. \citet{nog03dmlyr};
  %% AK Cnc
    104. \citet{men96akcnc};
    105. \citet{kat94akcnc};
  %% SS UMi
    106. \citet{kat02v344lyr};
    107. \citet{kat98ssumi};
    108. \citet{kat00ssumi};
  %% SX LMi
    109. \citet{nog97sxlmi};
    110. \citet{wag98sxlmi};
    111. \citet{kat01sxlmi};
  %% V701 Tau
    112. T. Kato et al., in preparation;
  %% V550 Cyg
    113. H. Iwamatsu et al., cf. vsnet-alert 5200; other periods
    are possible;
  %% V1504 Cyg
    114. \citet{pav02v1504cyg};
    115. \citet{nog97v1504cyg};
  %% KS UMa
    116. \citet{pat01SH};
    117. \citet{ole03ksuma};
    118. VSNET observations;
  %% RZ Sge
    119. \citet{sem97rzsge};
    120. \citet{kat96rzsge};
    121. \citet{bon82rzsge};
  %% V1208 Tau
    122. T. Kato et al., vsnet-alert 4165 and
    $<$http://www.kusastro.kyoto-u.ac.jp/vsnet/DNe/j0459.html$>$;
  %% IR Gem
    123. \citet{sha84irgem};
    124. \citet{kat01irgem};
  %% TY Psc
    125. \citet{kun01typsc};
    126. VSNET observations;
  %% CY UMa
    127. \citet{kat99cyuma};
    128. \citet{har95cyuma};
  %% PU Per
    129. \citet{kat99puper};
  %% FO And
    130. \citet{kat95foand};
  %% VW CrB
    131. \citet{nov97vwcrb};
    132. \citet{ant96vwcrb};
    133. VSNET observations;
  %% KV And
    134. \citet{kat95kvand};
    135. K. Tanabe et al., in preparation;
  %% AW Sge
    136. G. Masi et al., in preparation; other one-day alias is
    possible;
  %% NSV 10934
  %% V368 Peg
    137. J. Pietz, vsnet-alert 3317;
    138. \citet{ant98v368pegftcamv367pegv2209cyg};
    139. VSNET Collaboration data;
  %% OU Vir
    140. T. Kato et al., vsnet-campaign-dn 3656
    and VSNET observations;
    141. \citet{van00ouvir};
    142. \citet{szk02SDSSCVs};
  %% CC Cnc
    143. \citet{kat02cccnc};
    144. \citet{kat97cccnc};
  %% DM Dra
    145. \citet{kat02dmdra};
    146. \citet{ste82dmdra};
    147. T. Kato et al., vsnet-campaign-dn 3542;
  %% VZ Pyx
    148. \citet{kat97vzpyx};
    149. \citet{kiy99vzpyx};
  %% IY UMa
    150. \citet{uem00iyuma};
    151. \citet{pat00iyuma};
    152. M. Uemura et al., in preparation;
  %% AY Lyr
    153. \citet{uda88aylyr};
    154. \citet{szy87aylyr};
    155. \citet{pat79SH};
  %% V1251 Cyg
    156. \citet{kat95v1251cyg};
  %% HT Cas
    157. \citet{zha86htcas};
    158. \citet{wen85htcas};
    159. \citet{wen87htcas};
  %% QY Per
    160. \citet{kat00qyperiauc}; see also
    $<$http://www.kusastro.kyoto-u.ac.jp/vsnet/DNe/qyper.html$>$;
  %% QW Ser
    161. D. Nogami et al., in prepartion;
    162. \citet{kat99qwser};
    163. VSNET observations;
  %% BE Oct
    164. J. Kemp and J. Patterson, vsnet-obs 3461;
  %% VW Hyi
    165. \citet{vog74vwhyi};
    166. \citet{hae79lateSH};
    167. \citet{sch80vwhyi};
    168. \citet{lil96vwhyi};
  %% WX Hyi
    169. \citet{bai79wxhyiv436cen};
  %% Z Cha
    170. \citet{war88zcha};
    171. \citet{war74zcha};
    172. \citet{bai79zcha};
    173. \citet{vog82zcha};
    174. \citet{kuu91zcha};
  %% TT Boo
    175. \citet{kat95ttboo};
  %% WY Tri
    176. \citet{van01wytri};
    177. M. Uemura et al., vsnet-campaign-dn 376;
  %% RZ Leo
    178. \citet{ish01rzleo}; affected by beat phenomenon?;
  %% V630 Cyg
    179. \citet{nog01v630cyg};
  %% SU UMa
    180. \citet{uda90suuma};
  %% V1113 Cyg
    181. \citet{kat96v1113cyg};
    182. \citet{kat01v1113cyg};
  %% J1730
    183. M. Uemura et al., vsnet-campaign-dn 1786;
    184. T. Vanmunster et al., vsnet-campaign-dn 1792;
    185. VSNET observations;
  %% AW Gem
    186. \citet{kat96awgem};
  %% CU Vel
    187. N. Vogt (1981) Habilitation Thesis, Bochum University,
    see \citet{men96cuvel};
    188. Superhump periods may strongly vary.  T. Kato et al.
    (vsnet-campaign-dn 3136) gives a mean period of 0.08085(3) d from
    the 2002 December observation;
  %% DH Aql
    189. M. Uemura, vsnet-campaign-dn 2699 and VSNET Collaboration
    data;
    190. \citet{nog95dhaql};
  %% PV Per
    191. \citet{van97pvper};
    192. VSNET observations;
  %% HS Vir
    193. \citet{kat98hsvir};
    194. \citet{kat01hsvir};
  %% V359 Cen
    195. \citet{kat02v359cen};
  %% V660 Her
    196. J. Pietz, vsnet-alert 3162; the correct alias selected
    based on the orbital period in \citet{tho03kxaqlftcampucmav660herdmlyr};
    197. \citet{spo98alcomv544herv660herv516cygdxand};
    198. VSNET observations;
  %% V503 Cyg
    199. \citet{har95v503cyg};
    200. \citet{spo00v503cyg};
    201. \citet{kat02v503cyg};
  %% BR Lup
    202. \citet{odo87brlup};
    203. \citet{men98brlup};
    204. vsnet-campaign-dn 3584, T. Kato et al., in preparation;
  %% NSV 09923
    205. T. Kato et al., vsnet-campaign-dn 3821;
  %% RX Cha
    206. \citet{kat01rxcha};
    207. VSNET observations;
  %% V877 Ara
  %% AB Nor
  %% CP Dra
    208. R. Ishioka et al., in preparation, cf. vsnet-campaign-dn 556;
    209. \citet{kol79cpdraciuma};
  %% V369 Peg
    210. \citet{kat01v369peg};
  %% TU Crt
    211. \citet{men98tucrt};
    212. R. Ishioka et al., in preparation;
    213. \citet{wen93tucrt};
    214. \citet{haz93tucrt};
  %% HV Aur
    215. \citet{nog95hvaur};
    216. T. Kato et al., vsnet-campaign-dn 3061;
  %% V589 Her
    217. T. Vanmunster, vsnet-alert 7279; other possible period
    0.0950 d;
  %% EF Peg
    218. \citet{kat02efpeg};
    219. \citet{how93efpeg};
    220. VSNET observations;
    221. K. Matsumoto et al., in preparation;
  %% TY PsA
    222. \citet{bar82typsa};
    223. \citet{war89typsa};
    224. VSNET observations;
  %% BF Ara
    225. \citet{kat03bfara};
    226. \citet{kat01bfara};
    227. \citet{bru83bfara};
  %% KK Tel
  %% V452 Cas
    228. T. Vanmunster and B. Fried, vsnet-alert 3707;
    229. M. Uemura et al., vsnet-alert 3711;
  %% DV UMa
    230. \citet{pat00dvuma};
    231. M. Uemura et al., in preparation; see also
    $<$http://www.kusastro.kyoto-u.ac.jp/vsnet/DNe/dvuma9912.html$>$;
    232. \citet{nog01dvuma};
  %% V419 Lyr
    233. \citet{nog98gxcasv419lyr};
    234. M. Uemura et al., in preparation; see also vsnet-alert 3401;
  %% UV Gem
    235. T. Vanmunster, vsnet-alert 7284;
    236. \citet{kat01uvgemfsandaspsc};
    237. VSNET observations;
  %% V344 Lyr
    238. \citet{kat93v344lyr};
  %% YZ Cnc
    239. \citet{vanpar94suumayzcnc};
    240. \citet{kat01yzcnc};
    241. \citet{mof74yzcnc};
    242. \citet{szk84AAVSO};
  %% GX Cas
    243. VSNET observations;
    244. M. Uemura et al., in preparation;
  %% V725 Aql
    245. \citet{uem01v725aql};
    246. \citet{haz96v725aql};
    247. \citet{nog95v725aql}; the terminal portion of the
    superoutburst was observed; a large-amplitude superhump-like signal
    likely indicated a regrowth of the superhumps;
  %% MN Dra
    248. \citet{ant02var73dra};
    249. \citet{nog03var73dra}; the superhump period listed in the
    table is an average of the two different superoutbursts in
    \citet{nog03var73dra};
  %% NY Ser
    250. \citet{nog98nyser};
  %% TU Men
    251. \citet{StolzSchoembs};
    252. \citet{sto81tumen1};
    253. \citet{sto81tumen2};
    254. VSNET observations.  Outbursts with intermediate lengths
    are known; a typical recurrence time of long outbursts is given.
}
\end{table*}

\subsection{Distribution of Superhump Periods}

\begin{figure}
  \begin{center}
%    \FigureFile(88mm,60mm){pshhis.eps}
    \FigureFile(88mm,60mm){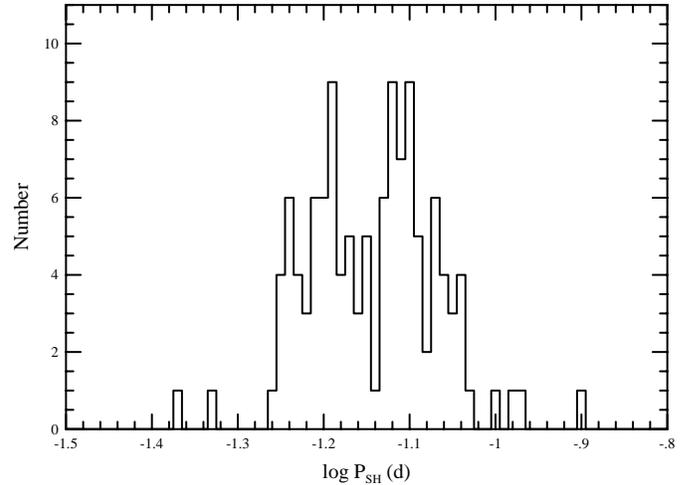}
  \end{center}
  \caption{Distribution of superhump periods ($P_{\rm SH}$) in SU UMa-type
  dwarf novae.  The data are taken from table \ref{tab:regrowth}.}
  \label{fig:pshhis}
\end{figure}

   Since table \ref{tab:regrowth} presents the most up-to-date and complete
statististics of SU UMa-type dwarf novae, we first reexamined the basic
statistics of SU UMa-type dwarf novae.  Figure \ref{fig:pshhis} shows
a distribution of the superhump periods ($P_{\rm SH}$) based on this
new material.  This figure supersedes the previously published statistics
(e.g. Fig. 6 of \cite{kol99CVperiodminimum}).  Even with the new samples,
a discrepancy still exists between the observed and theoretically predicted
distributions.  This new statistics confirms the claimed deficiency of
the shortest period systems (\cite{kol99CVperiodminimum};
\cite{bar03CVminimumperiod}).

\subsection{Distribution of Supercycles}

   We reexamined the distribution of supercycles, whose importance has been
discussed by a number of authors (\cite{hel01eruma}; \cite{nog98CVevolution};
\cite{war95suuma}).  From the standpoint of the disk-instability theory,
this distribution is considered to reflect the distribution of mass-transfer
rates \citep{ich94cycle}.

\begin{figure}
  \begin{center}
%    \FigureFile(88mm,60mm){his.eps}
    \FigureFile(88mm,60mm){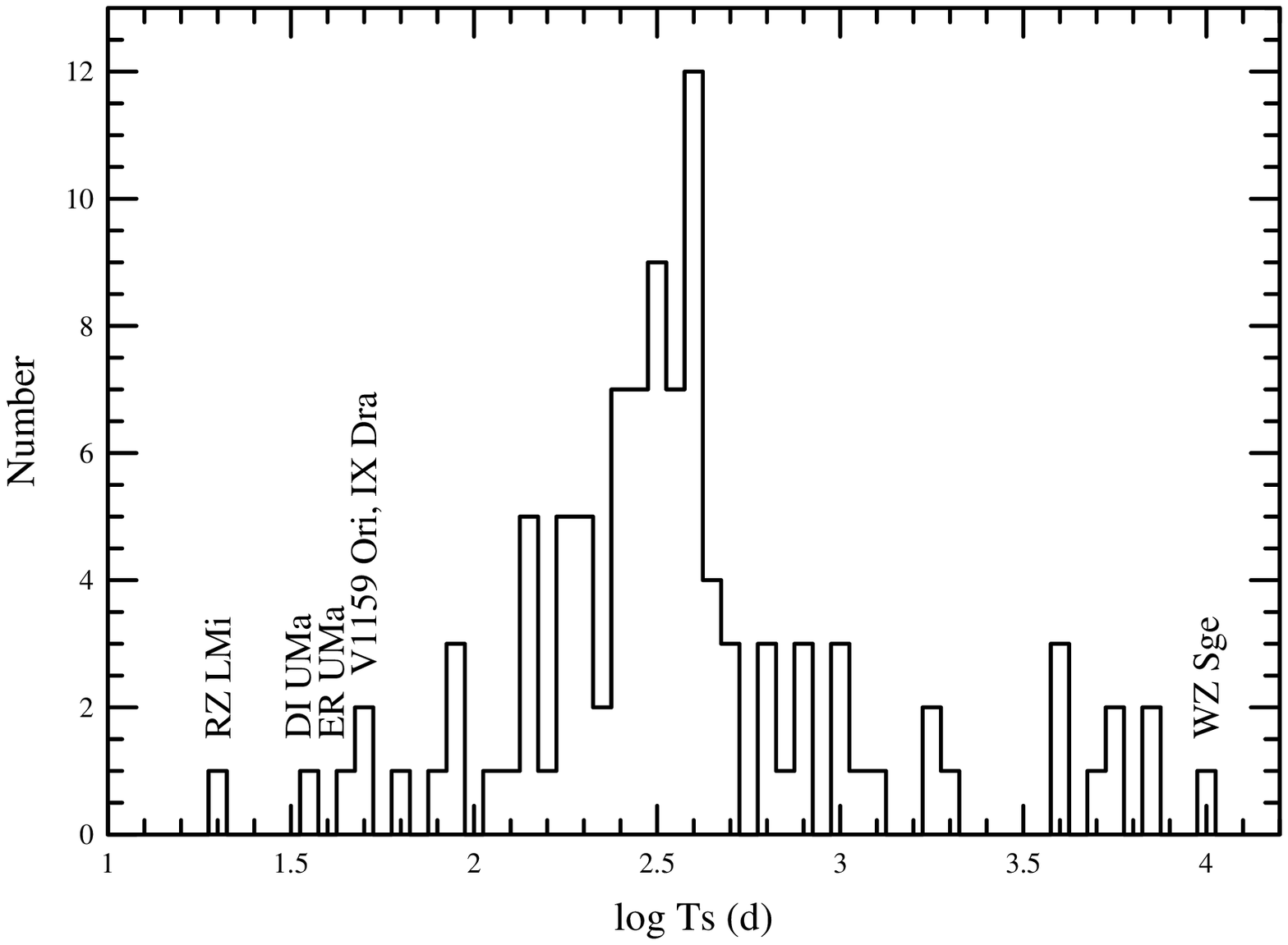}
  \end{center}
  \caption{Distribution of supercycle lengths ($T_s$) in SU UMa-type dwarf
  novae.  The data are taken from table \ref{tab:regrowth}.}
  \label{fig:his}
\end{figure}

   Figure \ref{fig:his} shows the resultant distribution of supercycle
lengths in SU UMa-type dwarf novae.  When there are different possibilities
of supercycle lengths are present, we chose the most likely period.
When a range of supercycle lengths is presented, we took a logarithmic mean
of the extreme values.  When only a lower limit of supercycle lengths is
available, we adopted the lower limit as a representative of the supercycle
lengths.  This figure supersedes Fig. 1 in \citet{hel01eruma}, who used
\cite{war95suuma}.  The increase of the number of samples since 1995
is remarkable.  Most of SU UMa-type dwarf novae have 2.0$< \log T_s <$3.0.
The objects below $\log T_s <$1.8 are ER UMa stars, except for Var 73 Dra
($\log T_s \sim$1.78: \citet{nog03var73dra}), whose relation with the
ER UMa stars is still unclear.
Despite intensive efforts to search for transitional objects between
ER UMa stars and usual SU UMa stars, any attempt has not yet been
successful \citep{kat03bfara}.  Many of the objects with $\log T_s >$3.5
are WZ Sge-type stars \citep{kat01hvvir}.  Although there have been
a suggestion that SU UMa-type dwarf novae and WZ Sge-type stars comprise
a continuous entity (e.g. \cite{pat96alcom}), the most up-to-data seem
to illustrate a gap around $\log T_s \sim$3.5.  The reality of the gap
needs to be verified by future studies.

\begin{figure}
  \begin{center}
%    \FigureFile(88mm,60mm){pshts.eps}
    \FigureFile(88mm,60mm){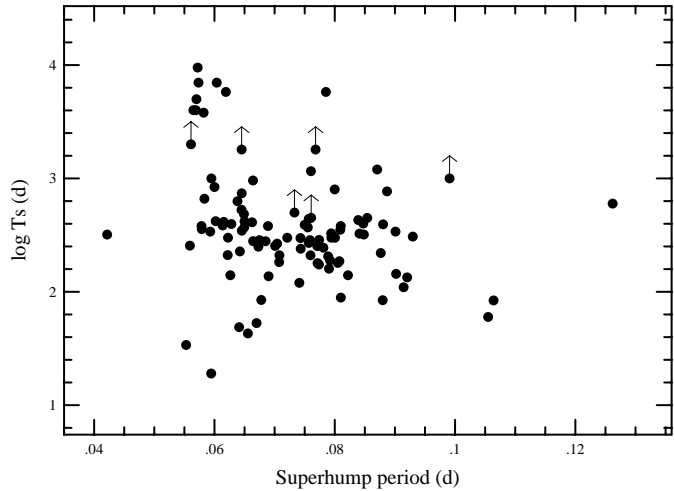}
  \end{center}
  \caption{Relation between $P_{\rm SH}$ and $T_s$.
  The data are taken from table \ref{tab:regrowth}.  The arrows represent
  lower limits of $T_s$'s.}
  \label{fig:pshts}
\end{figure}

   Figure \ref{fig:pshts} shows the relation between $P_{\rm SH}$ and $T_s$.
Although there is a tendency of a slight increase of typical $T_s$ toward
shorter $P_{\rm SH}$, the new statistics revealed that many of SU UMa-type
dwarf novae have a fairly typical value of $\log T_s \sim$2.5
($T_s \sim$300 d) regardless of the superhump period (or orbital period).
There is, however, a wide-spread distribution of $T_s$ around the period
0.055$\leq P_{\rm SH} \leq$0.065 d.  The systems with short supercycle
lengths ($\log T_s \leq$1.8) are ER UMa stars, and those with long
supercycle lengths ($\log T_s \geq$3.6) are WZ Sge-type stars.

   Although a concentration of these objects around the shortest $P_{\rm SH}$
has been widely discussed (cf. \cite{nog98CVevolution};
\cite{war98CVreviewWyo}; \cite{hel01eruma}), the present study more clearly
illustrated the segregation between these populations.  A previously
proposed picture of a continuous entity between usual SU UMa-type dwarf
novae and WZ Sge stars (cf. \cite{nog96cthya}; \cite{bab00v1028cyg})
seems to have become less concrete.  There are also a small number of objects
with long $T_s$ and long $P_{\rm SH}$, whose existence was not clear in
the past studies.  One of these objects (RZ Leo) has been studied
thoroughly (\cite{men02CVBD}; \cite{ish01rzleo}), and has been shown to
have a normal lower main-sequence secondary, rather than a brown dwarf.
The expected low mass-transfer rate in such a system should therefore be
understood as a result of a wide variety of systems even below the period
gap.  On the other hand, extremely short $T_s$ systems seem to be
concentrated in the short $P_{\rm SH}$ range, supporting the earlier
finding.

\subsection{Brightening near Termination of Superoutburst Plateau}

   We first inspected the objects with multiple superoutburst observations
in order to check the possible variation of the brightening feature near
the termination of a superoutburst plateau.  Table \ref{tab:brightening}
shows the summary.  This result indicates that whether or not there is a
brightening feature near the termination of a superoutburst plateau
is primarily dependent on the object, rather than individual
superoutbursts.  DV UMa and QW Ser are the only examples among
the well-observed objects which probably showed different types of
superoutbursts.  These objects apparently need further full-superoutburst
observations on every possible occasions.

\begin{table}
\caption{Brightening near Termination of Superoutburst Plateau in Different
         Superoutbursts.}
\label{tab:brightening}
\begin{center}
\vspace{10pt}
\begin{tabular}{cccc}
\hline\hline
Object    & $P_{\rm SH}$ & \multicolumn{2}{c}{Number of superoutbursts} \\
          & & \multicolumn{2}{c}{with/without brightening} \\
          & & Yes & No \\
\hline
V592 Her & 0.056498 & 2 & 0 \\
WX Cet   & 0.05949  & 3 & 0 \\
KV Dra   & 0.060005 & 2 & 0 \\
OY Car   & 0.064544 & 2 & 0 \\
TV Crv   & 0.0650   & 2 & 0 \\
CT Hya   & 0.06643  & 0 & 4 \\
DM Lyr   & 0.0673   & 0 & 2 \\
RZ Sge   & 0.07042  & 0 & 3 \\
CY UMa   & 0.07210  & 0 & 2 \\
KV And   & 0.07434  & 0 & 2 \\
CC Cnc   & 0.075518 & 0 & 2 \\
IY UMa   & 0.07588  & 0 & 2 \\
AY Lyr   & 0.07597  & 0 & 2 \\
QW Ser   & 0.07709  & 1 & 1 \\
VW Hyi   & 0.07714  & 0 & 4 \\
Z Cha    & 0.07740  & 0 & 2 \\
EF Peg   & 0.08705  & 0 & 2 \\
DV UMa   & 0.08869  & 1 & 1 \\
GX Cas   & 0.09297  & 0 & 2 \\
V725 Aql & 0.09909  & 2 & 0 \\
\hline
\end{tabular}
\end{center}
\end{table}

\begin{figure}
  \begin{center}
%    \FigureFile(88mm,60mm){bright.eps}
    \FigureFile(88mm,60mm){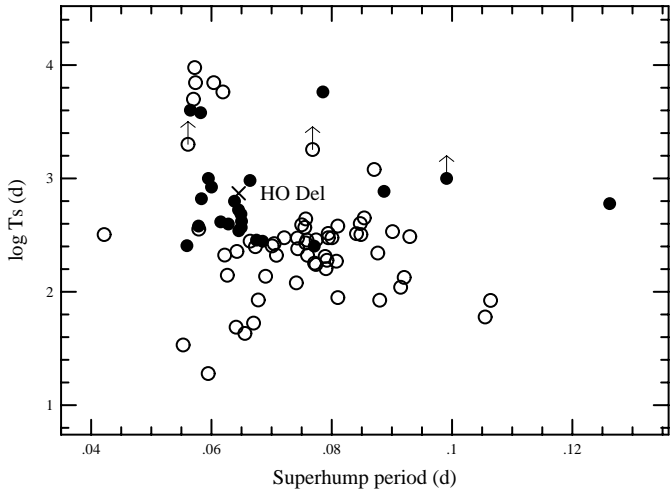}
  \end{center}
  \caption{Brightening near the termination of the superoutbursts.
  The open and filled circles represent no brightening and with brightening
  (No/No? and Yes/Yes? in table \ref{tab:regrowth}), respectively,
  including suspicious cases.  The arrows represent lower limits of
  $T_s$.  The location of HO Del (with brightening) is marked with
  a cross.}
  \label{fig:bright}
\end{figure}

   Figure \ref{fig:bright} shows a distribution of objects with or without
brightening near the termination of the superoutbursts.  The open and filled
circles represent no brightening and with brightening (No/No? and Yes/Yes?
in table \ref{tab:regrowth}), respectively, including suspicious cases.
(DV UMa and QW Ser were included as systems ``with brightening'').
There is a strong concentration of systems with brightening in short
$P_{\rm SH}$ and long $T_s$ systems, although WZ Sge-type stars usually
do not show this type of brightening.  Several long-$P_{\rm SH}$ systems
showing brightening are clearly confined to the long $T_s$ region.
The location of HO Del (with brightening) is consistent with the general
tendency.

\subsection{Regrowth of Superhumps}

   We inspected the objects with multiple superoutburst observations
in order to check the possible variation of the superhump regrowth near
the termination of a superoutburst plateau.  Table \ref{tab:regrowthind}
shows the summary.  The symbols are as in table \ref{tab:brightening}.
As in brightening feature near the termination of a superoutburst plateau,
the existence of the superhump regrowth is primarily dependent on the
object, rather than individual superoutbursts.
RZ Sge is a possible exception which showed both types of superoutbursts.

\begin{table}
\caption{Regrowth of Superhumps in Different Superoutbursts.}
\label{tab:regrowthind}
\begin{center}
\vspace{10pt}
\begin{tabular}{cccc}
\hline\hline
Object    & $P_{\rm SH}$ & \multicolumn{2}{c}{Number of superoutbursts} \\
          & & \multicolumn{2}{c}{with/without regrowth} \\
          & & Yes & No \\
\hline
SW UMa   & 0.05833 & 2 & 0 \\
WX Cet   & 0.05949 & 2 & 0 \\
TV Crv   & 0.0650  & 2 & 0 \\
CT Hya   & 0.06643 & 0 & 3 \\
RZ Sge   & 0.07042 & 1 & 2 \\
CY UMa   & 0.07210 & 0 & 2 \\
KV And   & 0.07434 & 0 & 2 \\
IY UMa   & 0.07588 & 0 & 2 \\
AY Lyr   & 0.07597 & 0 & 2 \\
VW Hyi   & 0.07714 & 0 & 4 \\
Z Cha    & 0.07740 & 0 & 2 \\
EF Peg   & 0.08705 & 0 & 2 \\
GX Cas   & 0.09297 & 0 & 2 \\
V725 Aql & 0.09909 & 2 & 0 \\
\hline
\end{tabular}
\end{center}
\end{table}

\begin{figure}
  \begin{center}
%    \FigureFile(88mm,60mm){growth.eps}
    \FigureFile(88mm,60mm){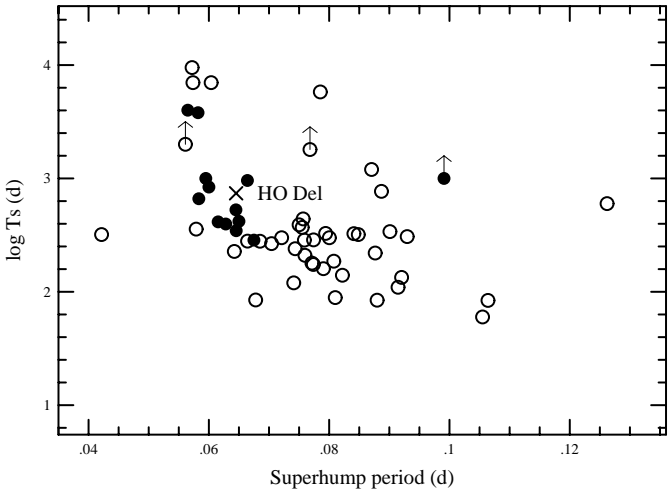}
  \end{center}
  \caption{Regrowth of superhumps near the termination of the superoutbursts.
  The open and filled circles represent no regrowth and with regrowth
  (No/No? and Yes/Yes? in table \ref{tab:regrowth}), respectively,
  including suspicious cases.  The arrows represent lower limits of
  $T_s$.  The location of HO Del (without regrowth) is marked with
  a cross.}
  \label{fig:growth}
\end{figure}

   Figure \ref{fig:growth} shows distribution of objects with or without
superhump regrowth near the termination of the superoutbursts.  The symbols
are the same as in figure \ref{fig:bright}.  A stronger preference of
short-$P_{\rm SH}$ systems, than in figure \ref{fig:bright}, is apparent,
although the location of the objects with superhump regrowth nearly
overlaps the location of the objects with brightening.  This figure
suggests that the appearance of superhump regrowth and brightening
near the termination of the superoutbursts is phenomenologically coupled.
This statistical finding may comprise another aspect of positive
correlations between brightening and superhump regrowth, which was claimed
by \citet{bab00v1028cyg} based on the analysis of the 1995 superoutburst
of V1028 Cyg.

   The location of HO Del is unusual in its apparent
absence of the superhump regrowth.  This deviation from the general
trend may be related to a rather low ($\sim$5.0 mag) outburst amplitude
of HO Del for a long $T_s$ object.  If this is the case, the conditions
necessary to manifest the unusual large-amplitude outbursts in some
rarely outbursting SU UMa-type dwarf novae, may be somehow responsible
for producing the late-time regrowth the superhumps.

   We do not discuss whether this argument can be observationally extended
to the most extreme WZ Sge-type dwarf novae, which may be a natural
extension of so-called TOADs, while recent theoretical interpretations
seem to prefer a different entity (\cite{osa02wzsgehump};
\cite{osa03DNoutburst}), in which a 1:2 resonance is claimed to be
essential for the manifestation of their unusual outbursts.
The exclusion of WZ Sge-type dwarf novae from figure \ref{fig:growth}
may be an artificial result our examination of these phenomena
(regrowth of superhumps and brightening)
restricted to pre-``dip'' observations of WZ Sge-type superoutbursts.
If we consider that the rebrightening stage of WZ Sge-type superoutbursts
actually corresponds to the late stage of SU UMa-type superoutbursts
discussed in this study, it may be possible both WZ Sge-type rebrightenings
and late stage brightening (and possibly regrowth of superhumps) have
the same underlying mechanism.  We leave this an observational open question,
partly because individual WZ Sge-type rebrightenings show a wide variety
[at least some of which bear more resemblance to normal outbursts
(e.g. \cite{kat97egcnc}; \cite{pat98egcnc}), while some of them
even look like a double superoutburst \citep{nog97alcom}], and partly
because observations are insufficient in most WZ Sge-type stars to judge
whether there was a regrowth of superhumps during such a stage.

   Although the original TOAD classification has been shown
to represent a rather loosely defined class of objects, the presently
discussed features may provide a better observational distinction for
SU UMa-type dwarf novae with unusual characteristics.

\section{Summary}

   We photometrically observed the 1994, 1996 and 2001 outbursts of
HO Del.  From the detection of secure superhumps, HO Del is confirmed to
be an SU UMa-type dwarf nova with a superhump period of 0.06453(6) d.
Based on the recent observations and the past records, the outbursts of
HO Del are found to be relatively rare, with the shortest intervals of
superoutbursts being $\sim$740 d.

   We also performed a literature survey of SU UMa-type dwarf novae, and
presented a new set of basic statistics.
This new statistics revealed that many of SU UMa-type dwarf novae have
a fairly typical value of supercycle lengths of $\sim$300 d regardless
of the superhump period.  There is, however, a wide-spread distribution
of $T_s$ around the period 0.055$\leq P_{\rm SH} \leq$0.065 d.
A previously proposed picture of a continuous entity between usual
SU UMa-type dwarf novae and WZ Sge stars seems to have become less concrete.

   The SU UMa-type dwarf novae with a brightening trend or with a regrowth
of superhumps near the termination of a superoutburst are found to be
rather tightly confined in a small region on the
(superhump period--supercycle length) plane.  These characteristics seem
to provide better criteria for a small, rather unusual population of
SU UMa-type dwarf novae, which likely correspond to, and would
better define, the objects previously selected by outburst amplitudes
(Tremendous Outburst Amplitude Dwarf Novae).
HO Del is rather unusual in that it is located in this region while
it did not show regrowth of superhumps.

\vskip 3mm

The authors are grateful to many observers who reported observations to
VSNET.  We are grateful to M. Moriyama, M. Reszelski, and J. Pietz for
their prompt announcements of the outburst detections of HO Del.
We are grateful to G. Masi, J. Pietz, B. Monard for providing their
observations of SU UMa-type dwarf novae prior to publication.
This work is partly supported by a grant-in-aid [13640239, 15037205 (TK),
14740131 (HY)] from the Japanese Ministry of Education, Culture, Sports,
Science and Technology.
This research has made use of the Digitized Sky Survey producted by STScI, 
the ESO Skycat tool, and the VizieR catalogue access tool.

\end{document}